\documentclass[conference]{IEEEtran}
\IEEEoverridecommandlockouts
\usepackage{cite}
\usepackage{amsmath,amssymb,amsfonts}
\usepackage{algorithmic}
\usepackage{graphicx}
\usepackage{textcomp}
\usepackage{subcaption} 
\usepackage{xcolor}
\usepackage{ mathrsfs }
\usepackage{amsmath}
\usepackage{amssymb}
\usepackage{siunitx}
\usepackage{multirow}
\newcommand{\linebreakand}{%
  \end{@IEEEauthorhalign}
  \hfill\mbox{}\par
  \mbox{}\hfill\begin{@IEEEauthorhalign}
}
\def\BibTeX{{\rm B\kern-.05em{\sc i\kern-.025em b}\kern-.08em
    T\kern-.1667em\lower.7ex\hbox{E}\kern-.125emX}}
\begin{document}

\title{Automatic Radar Signal Detection and FFT Estimation using Deep Learning
}

\author{
\IEEEauthorblockN{Akila Pemasiri}
\IEEEauthorblockA{\textit{Queensland University of Technology},\\
Australia \\
a.thondilege@qut.edu.au} 
\and
\IEEEauthorblockN{Zi Huang}
\IEEEauthorblockA{\textit{Queensland University of Technology},\\
Australia \\
z36.huang@hdr.qut.edu.au} 
\and
\IEEEauthorblockN{Fraser Williams}
\IEEEauthorblockA{\textit{Queensland University of Technology},\\
Australia \\
fraser.williams@hdr.qut.edu.au}
\and
\IEEEauthorblockN{Ethan Goan}
\IEEEauthorblockA{\textit{Queensland University of Technology},\\
Australia \\
ej.goan@qut.edu.au} 

\and
\IEEEauthorblockN{Simon Denman}
\IEEEauthorblockA{\textit{Queensland University of Technology},\\
Australia \\
s.denman@qut.edu.au} 
\and
\IEEEauthorblockN{Terrence Martin}
\IEEEauthorblockA{\textit{Revolution Aerospace}, Australia \\
terry@revn.aero} 

\and
\IEEEauthorblockN{Clinton Fookes}
\IEEEauthorblockA{\textit{Queensland University of Technology},\\
Australia \\
c.fookes@qut.edu.au}

}

\maketitle

\begin{abstract}
This paper addresses a critical preliminary step in radar signal processing: detecting the presence of a radar signal and robustly estimating its bandwidth. Existing methods which are largely statistical feature-based approaches face challenges in electronic warfare (EW) settings where prior information about signals is lacking. While alternate deep learning based methods focus on more challenging environments, they  primarily formulate this as a binary classification problem. In this research, we propose a novel methodology that not only detects the presence of a signal, but also localises it in the time domain and estimates its operating frequency band at that point in time. To achieve robust estimation, we introduce a compound loss function that leverages complementary information from both time-domain and frequency-domain representations. By integrating these approaches, we aim to improve the efficiency and accuracy of radar signal detection and parameter estimation, reducing both unnecessary resource consumption and human effort in downstream tasks.
\end{abstract}


\section{Introduction}
Detecting the presence of a radar signal and robustly estimating its bandwidth is a crucial preliminary step in radar signal processing. This reduces both the consumption of computing resources and human efforts in downstream tasks such as radar parameter estimation and derivation of specific signal characteristics for electronic support \cite{Richard2006}.

In low probability of intercept (LPI) signal detection, the primary objective remains the minimization of the detectable signal-to-noise ratio (SNR) \cite{Donoughue2019}. In an electronic warfare (EW) scenario, the receiver does not possess any prior information about the to-be-detected signals. Initial detection methods such as matched filters are not well-suited for scenarios where neither the received signal nor its modulation type are known \cite{Roman2000}. Furthermore, general likelihood ratio tests, which typically require prior knowledge about the modulation type of the received signal, are also not applicable in such cases \cite{Ly2017}.

Other approaches have primarily relied on statistical feature-based methodologies including energy detection \cite{Liang2008} and time-frequency domain detection \cite{Geroleo2012, Liu2015 }. However, these techniques face inherent limitations with degraded performance in challenging environments \cite{Zhang2023}. Recently, deep learning (DL)-based methods have received much attention for characterizing and classifying radar signals \cite{Shea2018,Huynh2021,Zi2023, Zi2024}.  DL-based signal detection methods have primarily been explored using two main approaches: time-frequency images \cite{Chen2019,Liu2018} (TFI), and  IQ sequences as model inputs \cite{Nuhoglu202,Su2021,Gao2019}. 

However, while DL-based methods have demonstrated superior results compared with the other methods, most existing DL-based methods are formulated as a binary classification task where the task is simply detecting the presence of a signal~\cite{Nuhoglu202,Su2021,Gao2019}. While some methods have sought to detect the signal and its location in the time domain \cite{Zhang2023}, in this paper we present a novel methodology to detect the signal, localise it in the time domain, and estimate the frequency content occupied by the signal at that point in time. To enable robust estimation of the signal duration and operating frequency band we utilise a compound loss function which is capable of learning complementary information shared between the time domain and frequency domain representations.




\section{Proposed Method}
\begin{figure}[htbp]
\centerline{\includegraphics[width=80mm]{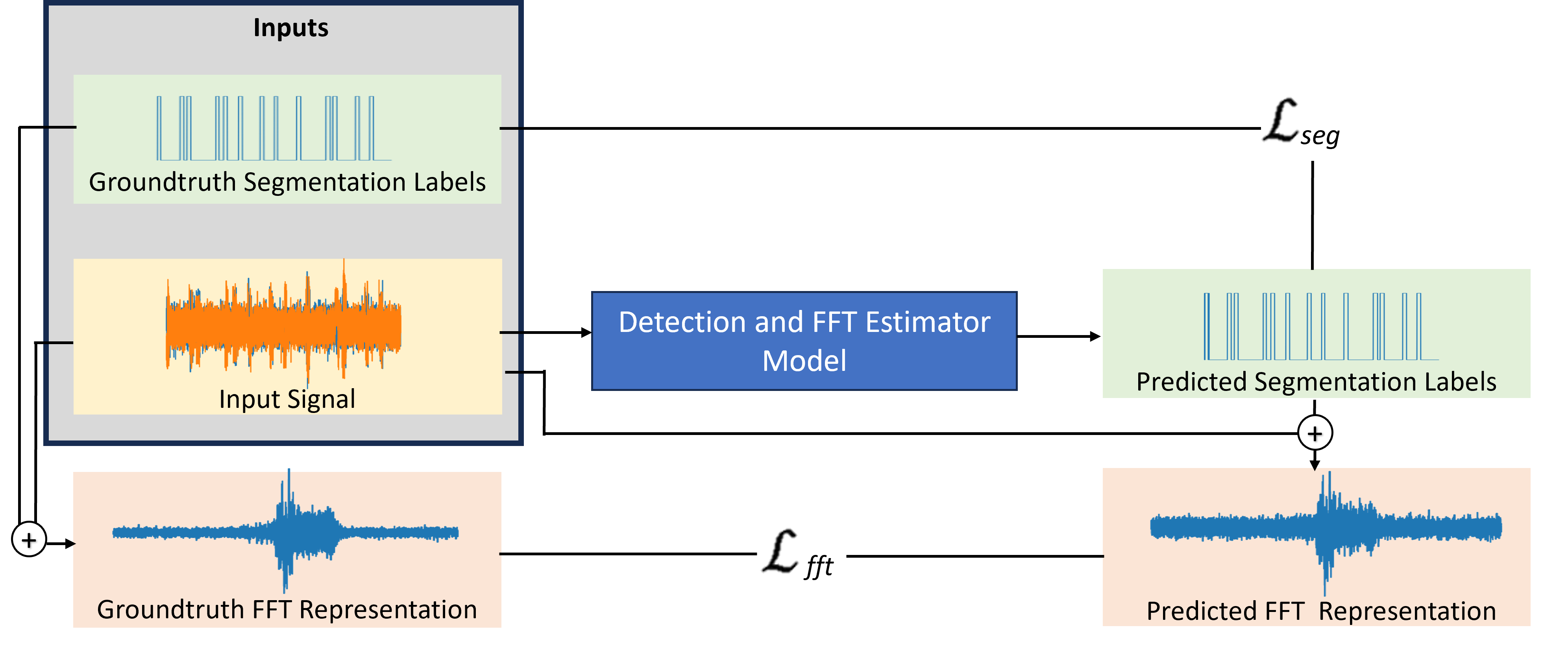}}
\caption{The training phase of the proposed method. The network receives as input the IQ signal and the groundtruth segmentation labels. Segmentation loss ($\mathcal{L}_{seg}$) and FFT loss ($\mathcal{L}_{fft}$) are calculated and used for the training of the model.} 
\label{overall_method_train}
\end{figure}

\begin{figure}[htbp]
\centerline{\includegraphics[width=80mm]{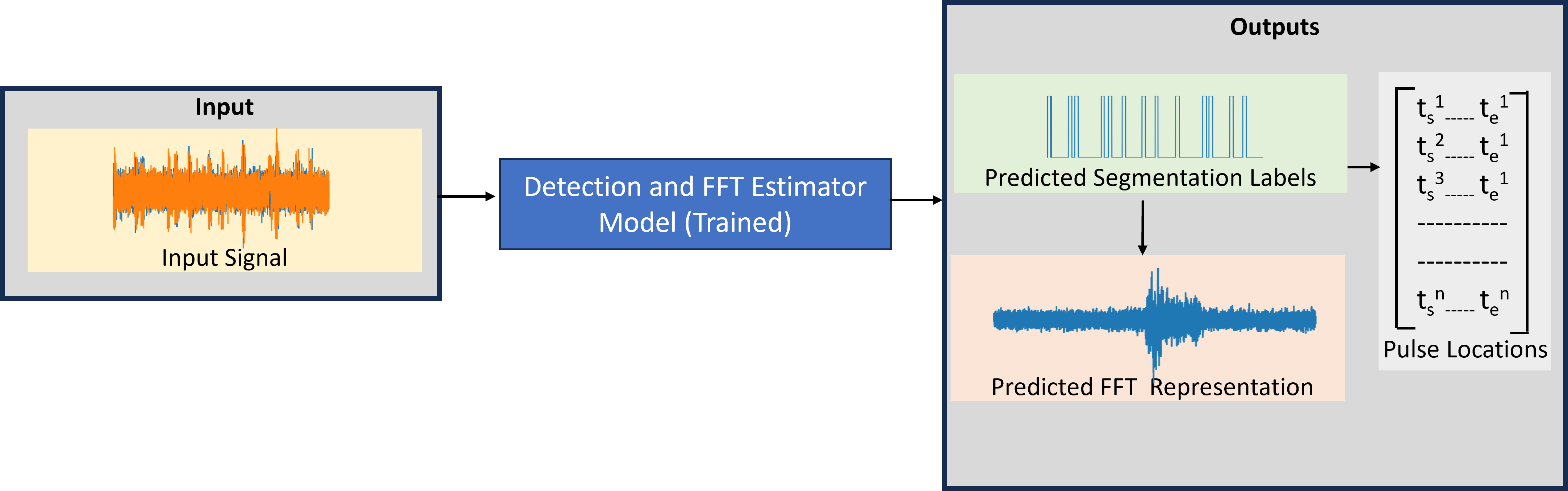}}
\caption{The inference phase of the proposed method. The network receives as input the IQ signal and estimates the pulse locations the FFT representation. For the pulse locations, $t_{s}^{n}$ denotes the starting time stamp of $n^{\text{th}}$ pulse and $t_{e}^{n}$ denotes the ending time stamp of the same pulse.}
\label{overall_method_test}
\end{figure}

The flow of data through  our model during the training phase is depicted in Fig. \ref{overall_method_train}, and information flow during the inference phase is depicted in Fig. \ref{overall_method_test}. During the training phase, the IQ signal and the corresponding groundtruth segmentation label are used as the input. In the groundtruth segmentation label, 1 denotes that a signal of interest is present at a point in time, and 0 denotes the absence of such a signal. Using the input signal and the groundtruth segmentation label, the groundtruth fast Fourier transform (FFT) representation is obtained \cite{Heckbert1995}. The ``Detection and FFT Estimator Model'' in Fig. \ref{overall_method_train} network model  estimates the prediction of the segmentation mask, which is then used with the input IQ signal to generate the  FFT representation for the signal of interest. 

In the inference phase (Fig. \ref{overall_method_test}) the input is the IQ sequence and the trained model will output the starting point and the end point of each pulse and the FFT representation of the incoming signal.

\begin{figure*}[hbtp]
\centerline{\includegraphics[width=120mm]{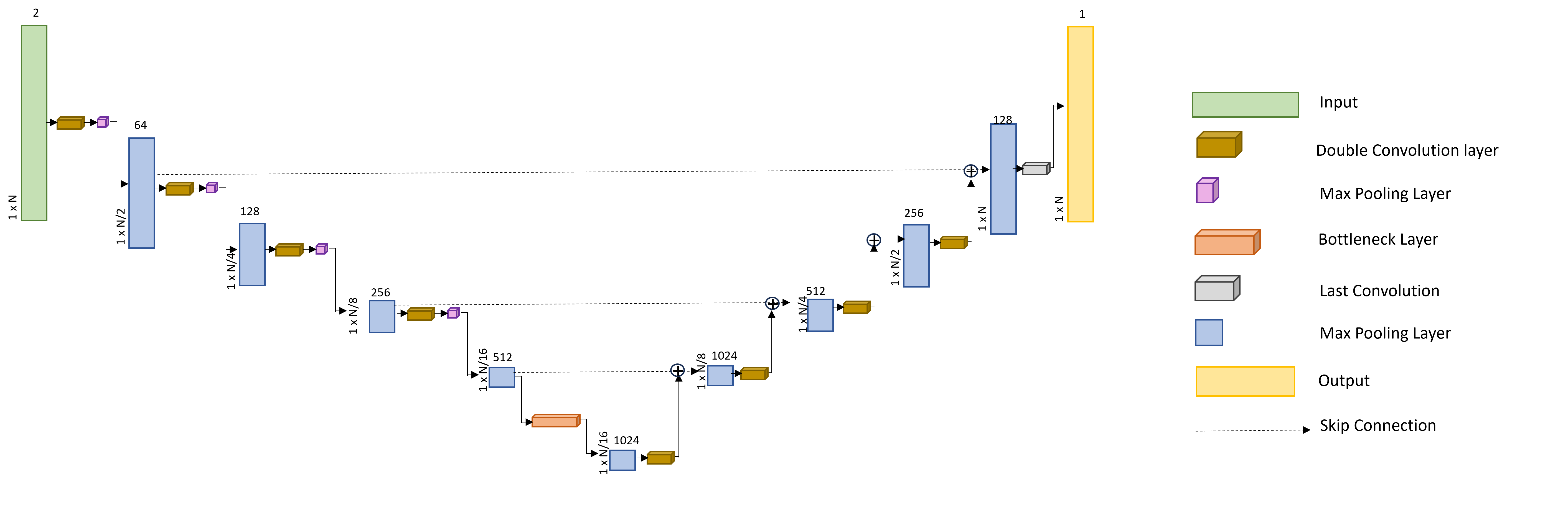}}
\caption{The Unet architecture we use in this work. $N$ is the sequence length under consideration. The details of the layers are discussed in Section \ref{network_archutecture} of this paper. }
\label{unet_architecture}
\end{figure*}

\subsection{Network Architecture}
\label{network_archutecture}

We use a UNet \cite{Ronneberger2015Unet} architecture for predicting the segmentation labels. This is because the standard UNet architecture modified to use 1D convolution operations has demonstrated superior performance compared to the other sequence segmentation methodologies on long and interleaved radar signals. We follow a similar architecture to \cite{Zi2024}. Fig. \ref{unet_architecture} depicts the network architecture we use in our work.

Both the contracting path and the expansive path of our network use Double Convolution Layers that consist of 2 blocks where each block contains a 1D  Convolution Layer followed by a 1D Batch Normalization layer and a Rectified Linear (ReLU) layer. To alleviate the information loss between down sampling and up sampling paths, skip connections are used and the features are concatenated across them.

We employ a compound loss ($L$) function which seeks to minimise both the segmentation loss and the FFT estimation loss \eqref{compundloss}, where $w_{\text{seg}}$ and $w_{\text{fft}}$ are the coefficients associated with the segmentation loss and FFT loss respectively. We use Binary Cross Entropy (BCE) loss \eqref{seg_loss} for the segmentation loss, and Mean Squared Error (MSE) loss \eqref{fft_loss} for the FFT estimation loss.

\begin{equation}
\mathcal{L} = w_{\text{seg}} \mathcal{L}_{\text{seg}} +w_{\text{fft}}\mathcal{L}_{\text{fft}}
\label{compundloss}
\end{equation}

\begin{equation}
\begin{aligned}
\mathcal{L}_{\text{seg}} & = \frac{1}{M} \sum_{i= 1}^M - \left( {\text{seg}_{\text{gt}}}_{i} \ast log\left(p_{i} \right) \right) \\
& + \left(1-{\text{seg}_{\text{gt}}}_{i} \ast log\left(1-p_{i} \right) \right)
\end{aligned}
\label{seg_loss}
\end{equation}

\begin{equation}
\mathcal{L}_{\text{fft}} = \frac{1}{M} \sum_{i= 1}^M  \left( {\text{gt}_{\text{fft}}}_{i} - {\text{pred}_{\text{fft}}}_{i} \right)
\label{fft_loss}
\end{equation}

In \eqref{seg_loss} and \eqref{fft_loss}, $M$ is the number of samples under consideration. In \eqref{seg_loss}, ${\text{seg}_{\text{gt}}}_{i}$ refers to the $i^{\text{th}}$  groundtruth segmentation label and $p_{i}$ is the estimated  probability of class 1. In \eqref{fft_loss},  ${\text{gt}_{\text{fft}}}_{i} $ refers to the $i^{\text{th}}$ normalized groundtruth FFT value and  ${\text{pred}_{\text{fft}}}_{i} $ refers to the $i^{\text{th}}$ normalized predicted FFT value.  

We train our models with a constant learning rate of $10^{-4}$ using the Adam optimizer\cite{adamOptimizer}. To reduce the number of experimental permutations, we set $w_{\text{seg}}$ and $w_{\text{fft}}$ in \eqref{compundloss} to 1. 

\section{Experiments}
\subsection{Dataset}
\begin{figure}[htbp]
    \centering
    \begin{subfigure}[b]{0.5\textwidth}
        \centering
        \includegraphics[height=2cm,width = 6cm]{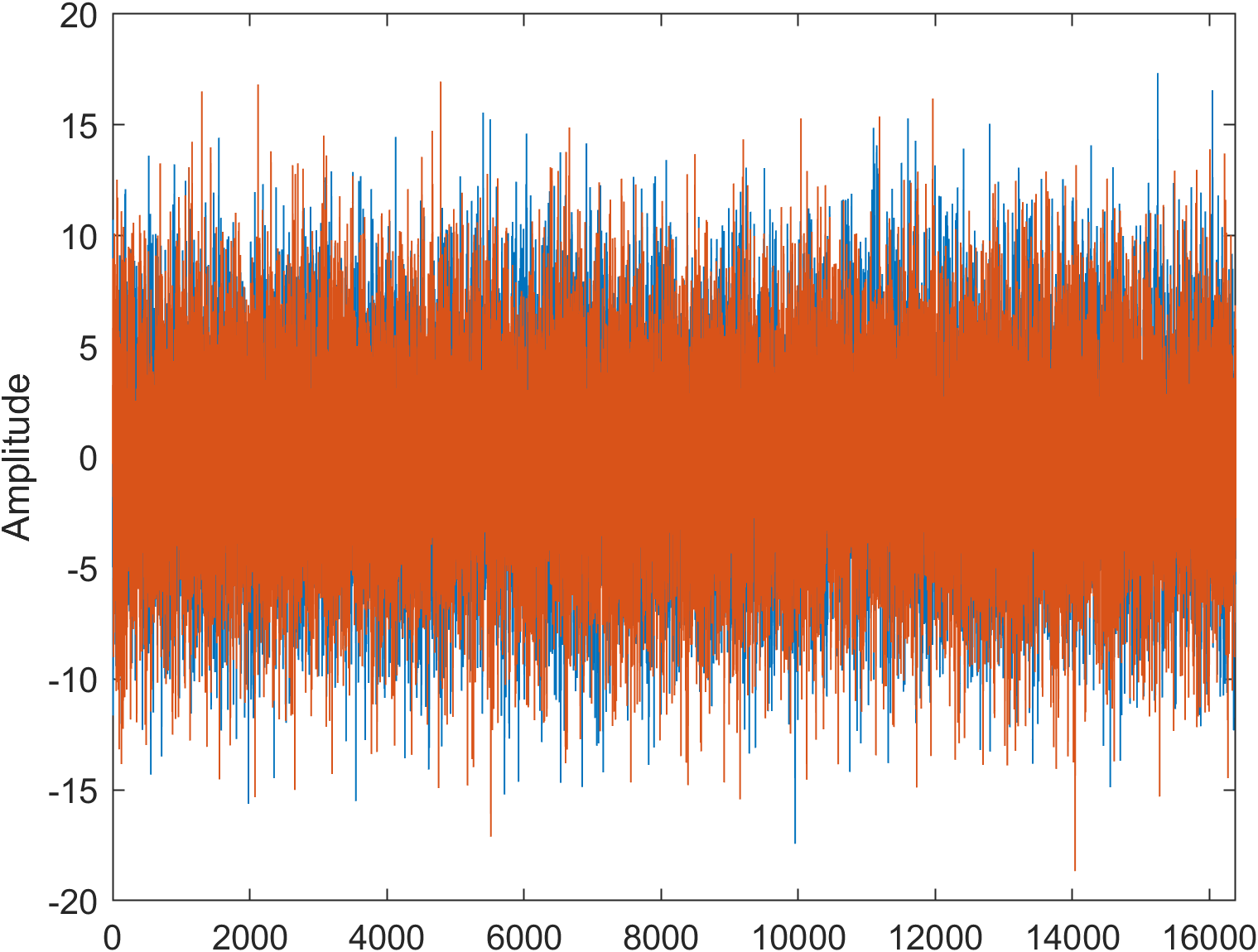}
        \caption{Raw IQ Signal.}
        \label{fig:sub1}
    \end{subfigure}
    \hfill
    \begin{subfigure}[b]{0.5\textwidth}
        \centering
        \includegraphics[height=3cm, ,width = 5.8cm]{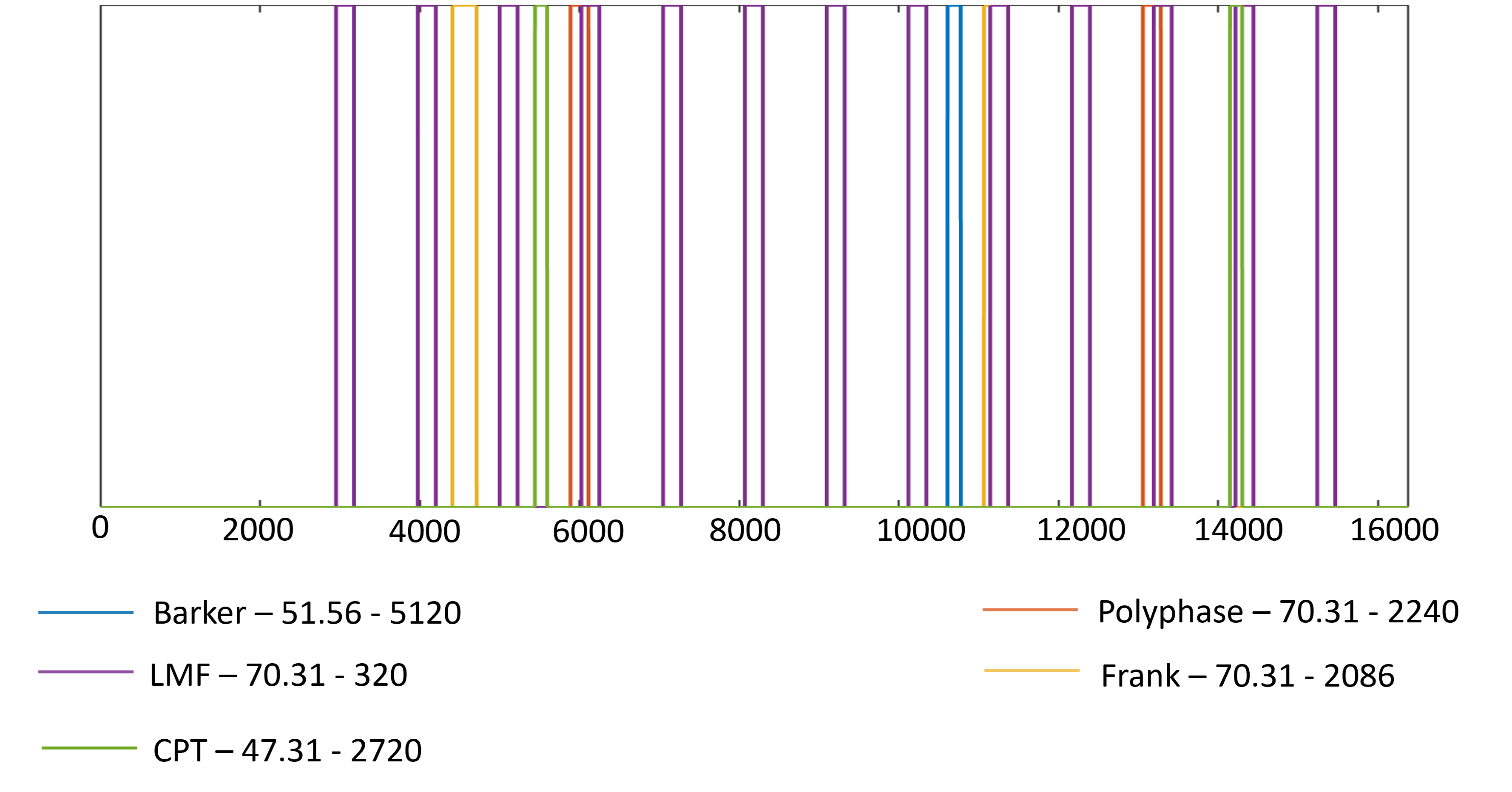}
        \caption{Presence of each signal. In the legend Barker - 10 - 100, refers to a signal with modulation type ``Barker'', with ``$10\si{\micro\second}$'' PW and `$10\si{\micro\second}$'' PRI.}
    \end{subfigure}
    \hfill
    \begin{subfigure}[b]{0.5\textwidth}
        \centering
        \includegraphics[height=2cm, ,width = 5.5cm]{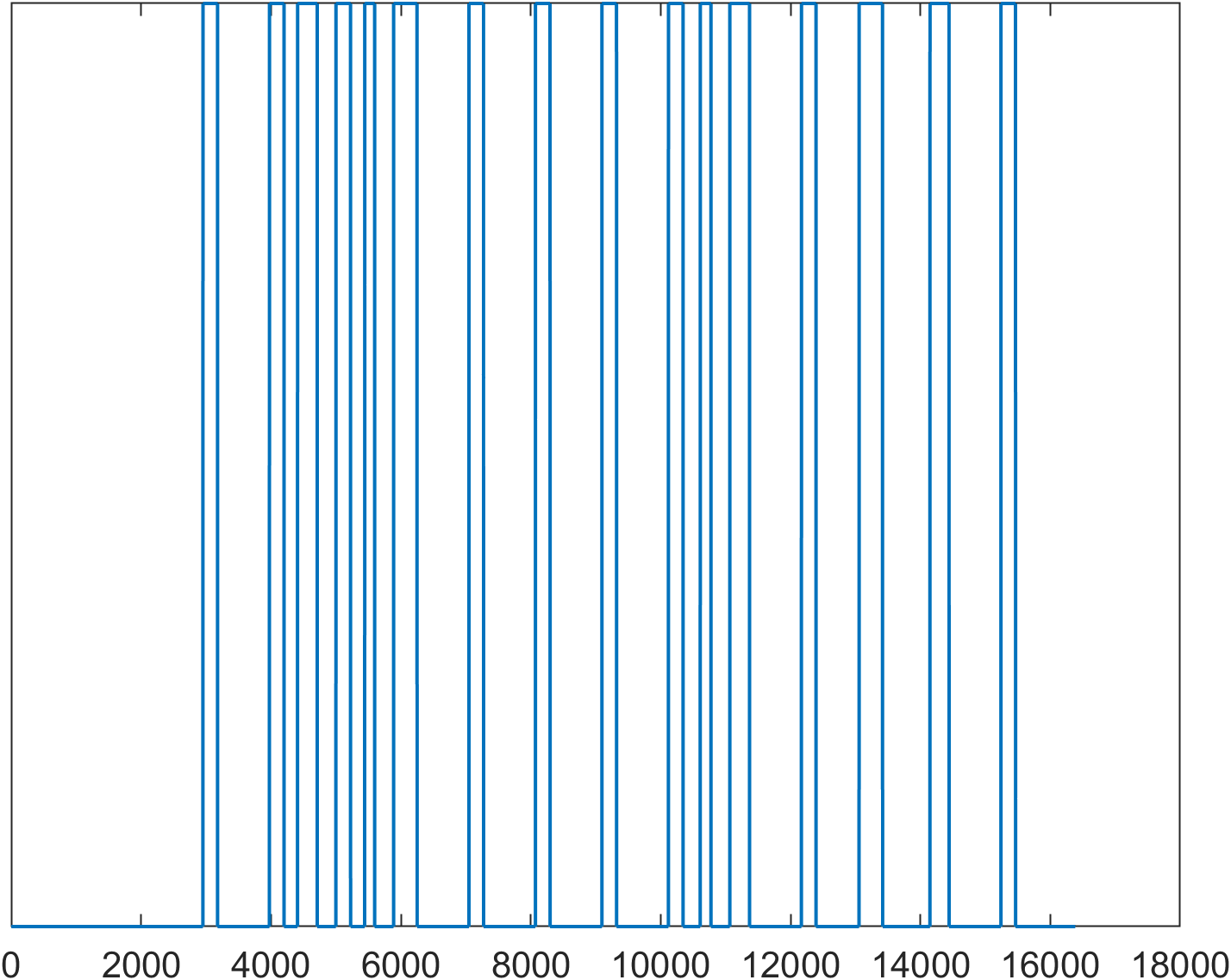}
        \caption{The annotation labels used in training the models.}
        \label{fig:sub3}
    \end{subfigure}
    \caption{A sample data sequence, with -17 dB SNR.}
    \label{dataset_figure}
\end{figure}

We use a new radar signal detection dataset which builds upon the datasets proposed in \cite{Zi2023, Zi2024}. We consider 5 radar signal classes: 1)  coherent unmodulated pulses (CPT), 2) Barker codes, 3) polyphase Barker codes, 4) Frank codes, and 5) Linear frequency-modulated (LFM) pulses. We consider the same parameter ranges for the parameters of pulse width (PW) and pulse repetition interval (PRI). In addition to the Additive white Gaussian noise (AWGN), in this dataset, we introduce channel impairments by incorporating frequency offsets and phase offsets. The frequency offsets of $20,000$, $40,000$, $60,000$, $80,000$ and $100,000$ Hz, and phase offsets from 0 to $\pi/2$ radians at $\pi/16$ intervals are used. We allow radar signals to freely alternate, capturing the dynamic nature of an electronic warfare setting \cite{Ge2019}. A sample data sequence from our dataset is depicted in Fig. \ref{dataset_figure}.
\section{Evaluation Protocol}
We use the train split of our dataset to train the models and use the independent test split to test the models. The training set contains 120, 000 signals, while the validation and test set each contain 20, 000 signals. 

We use the average F1 (\eqref{f1_score}, \eqref{precision}, \eqref{recall}) value over all the datapoints to evaluate the accuracy of the segmentation predictions.  In \eqref{precision} and  \eqref{recall}, $TP$, $FP$ and $FN$ refer to True Positives, False Positives and False Negatives respectively. 
\begin{equation}
 F_{1} = 2 \ast \frac{\text{Precision} \cdot \text{Recall}}{ \text{Precision} + \text{Recall} }
\label{f1_score}
\end{equation}

\begin{equation}
 Precision = \frac{\text{TP}}{\text{TP} + \text{FP}}
\label{precision}
\end{equation}

\begin{equation}
 Recall = \frac{\text{TP}}{\text{TP} + \text{FN}}
\label{recall}
\end{equation}

To calculate the similarity between the groundtruth FFT representation and the predicted FFT representation, we use cosine similarity between the two representations \eqref{cosine}. 

\begin{equation}
 \text{similarity} = \frac{\text{gt}_{\text{fft}} \cdot \text{pred}_{\text{fft}}}{ \left\| \text{gt}_{\text{fft}}\right \|+ \left\| \text{pred}_{\text{fft}}\right \|}
\label{cosine}
\end{equation}

To compare our method with the energy detection approach of \cite{Donoughue2019}, we use probability of correct detection ($P_{\text{d}}$)  and the probability of false detection($P_{\text{fa}}$).

 \section{Performance Evaluation}

We conduct our evaluations with the aims of: 1) Evaluating the effect of sequence length and the performance; 2) Evaluating    the contribution of the components in Equation \eqref{compundloss} to the overall performance; and 3) Comparing method with Energy detection \cite{Donoughue2019} method. 

Table \ref{ablative_seq_length} provides a summary of the results of experiments conducted on datasets with different sequence lengths and with different settings of the loss function shown in Equation \eqref{compundloss}. It can be observed that the proposed approach has performed consistently across different sequence lengths. Compared with the high SNR levels, the performance at the low SNRs remains low, and this is an expected trend and is consistent with similar radio recognition tasks \cite{Zi2023, Zi2024,  Jagannath}. From Table \ref{ablative_seq_length}, it can be observed that a significant performance gain can be obtained on the segmentation task by incorporating the $L_{fft}$ within the loss function. 

\begin{figure*}
    \centering
    \begin{subfigure}[b]{0.48\textwidth}
        \centering
        \includegraphics[width=0.6\textwidth]{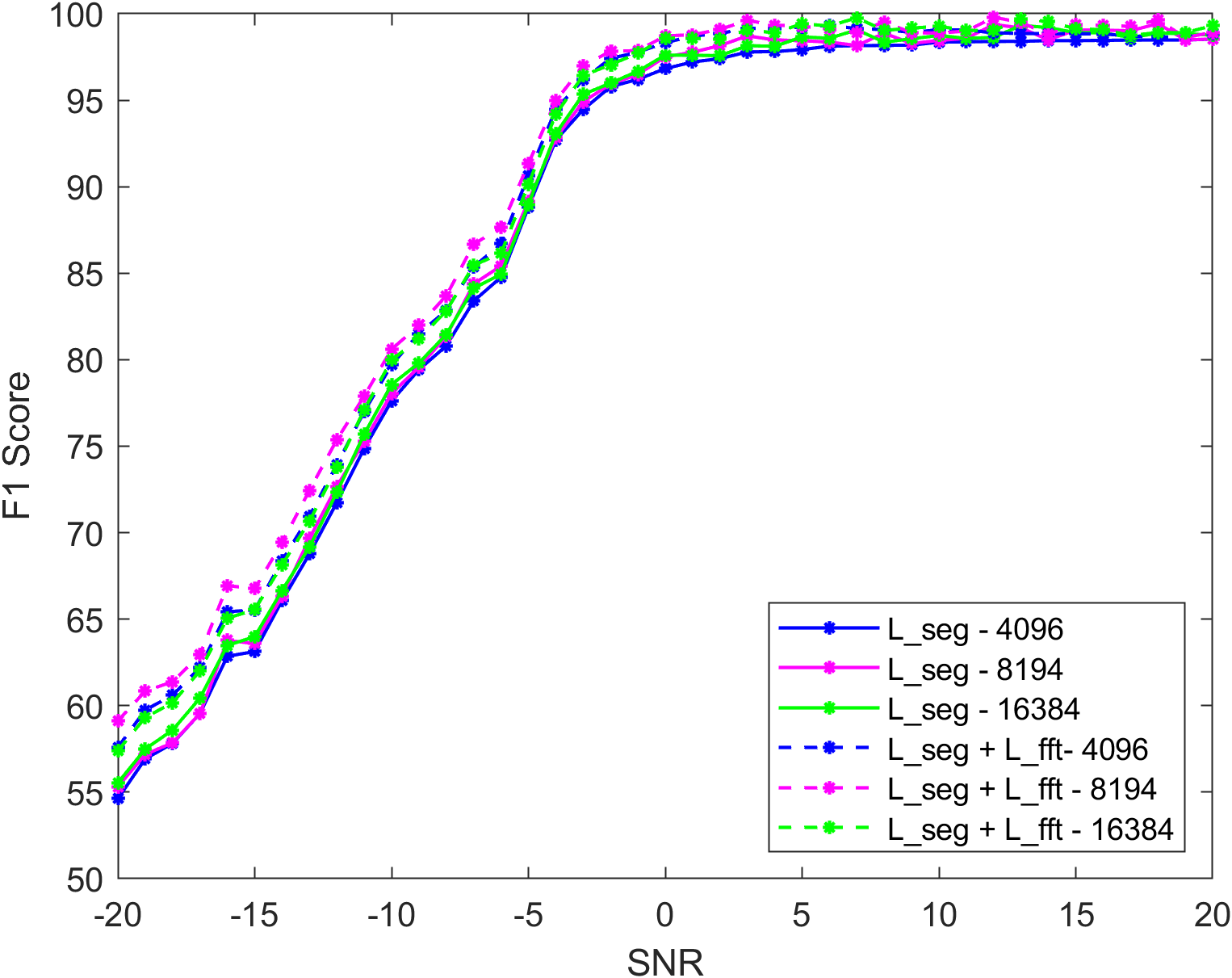}
        \caption{$F_{1}$ measure for segmentation prediction across  all the considered SNRs.}
        \label{performance_seg}
    \end{subfigure}
    \hfill
    \begin{subfigure}[b]{0.48\textwidth}
        \centering
        \includegraphics[width=0.6\textwidth]{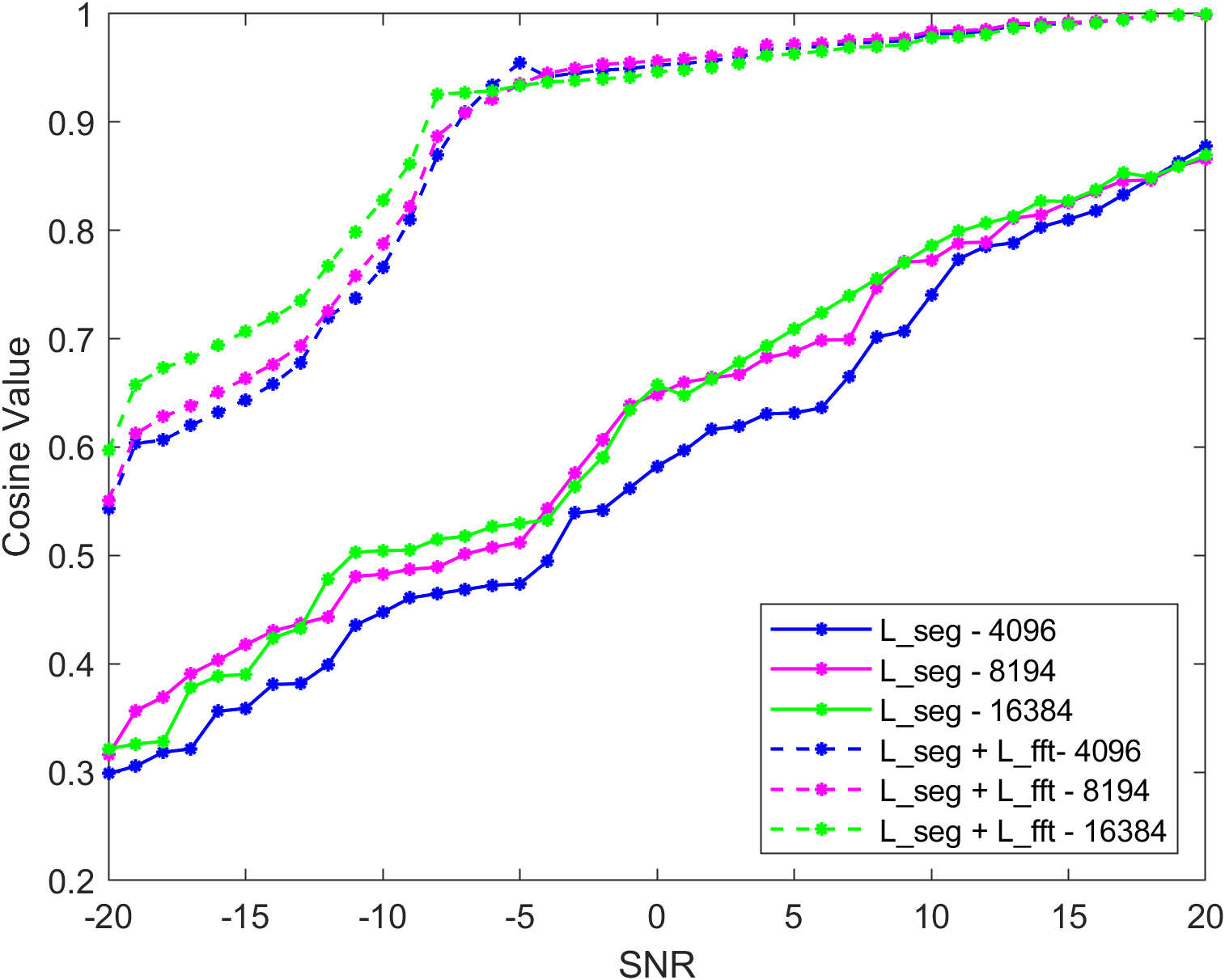}
        \caption{Cosine similarity for FFT prediction across  all the considered SNRs.}
        \label{performance_FFT}
    \end{subfigure}
    \caption{Performance evaluations for segmentation prediction and FFT prediction.}
    \label{fig_performence}
    \end{figure*} 
\begin{table}[]
\caption{Comparison of segmentation models on different
sequence lengths and different loss components based on $F_{1}$ score.}

\begin{tabular}{|l|l|l|l|l|l|l|}
\hline
\multicolumn{1}{|c|}{\begin{tabular}[c]{@{}c@{}}Loss Function\\ Components\end{tabular}} & \multicolumn{1}{c|}{\begin{tabular}[c]{@{}c@{}}Sequence \\ Length\end{tabular}} & \multicolumn{1}{c|}{-20dB} & \multicolumn{1}{c|}{-10dB} & \multicolumn{1}{c|}{0dB} & \multicolumn{1}{c|}{10dB} & \multicolumn{1}{c|}{20dB} \\ \hline
\multirow{3}{*}{$L_{seg}$} & 4096 & 54.63 & 77.62 & 96.83 & 98.35 & 98.53 \\ \cline{2-7} 
 & 8192 & 55.31 & 78.05 & 97.513 & 98.46 & 98.54 \\ \cline{2-7} 
 & 16384 & 55.54 & 78.56 & 97.61& 98.74 & 98.75 \\ \hline
\multirow{3}{*}{$L_{seg}$+$L_{fft}$} & 4096 & 57.59 & 79.70 & 98.33 & 99.07 & 98.87 \\ \cline{2-7} 
 & 8192 &59.13 & 80.60 & 98.72 & 98.88 & 98.92 \\ \cline{2-7} 
 & 16384 & 57.40 & 79.98 & 98.60 & 99.27 & 99.31\\ \hline
\end{tabular}
\label{ablative_seq_length}
\end{table}

Similarly, Table \ref{ablative_loss_components} provides a summary of the results of FFT estimation accuracy. Similar to the segmentation tasks, it can be observed the models have performed consistently across different sequence lengths. When compared with the models that only use $\mathcal{L}_{seg}$ in \eqref{compundloss}, it can be seen the models that use the compound loss function have obtained significant accuracy in FFT estimation even at the lower SNRs. Fig. \ref{fig_performence} depicts the performance on segmentation and FFT estimation, which highlights the impact of the compound loss function. Fig \ref{performance_FFT} highlights the significance of enabling the $\mathcal{L}_{fft}$ in the loss function. From Table \ref{ablative_seq_length} and Table \ref{ablative_loss_components}, it can be seen that employing the compound loss function has a positive effect on both segmentation and FFT estimation as the model is effectively learning the complementary information.

\begin{table}[]
\caption{Comparison of FFT estimation on different sequence lengths and different loss components. This records cosine similarity computed at -20, -10, 0, 10 and 20 dB SNR. }
\begin{tabular}{|l|l|l|l|l|l|l|}
\hline
\multicolumn{1}{|c|}{\begin{tabular}[c]{@{}c@{}}Loss Function\\ Components\end{tabular}} & \multicolumn{1}{c|}{\begin{tabular}[c]{@{}c@{}}Sequence \\ Length\end{tabular}} & \multicolumn{1}{c|}{-20} & \multicolumn{1}{c|}{-10} & \multicolumn{1}{c|}{0} & \multicolumn{1}{c|}{10} & \multicolumn{1}{c|}{20} \\ \hline
\multirow{3}{*}{$L_{seg}$} & 4096 & 0.299 & 0.447 & 0.582 & 0.740 & 0.878 \\ \cline{2-7} 
 & 8192 & 0.316 & 0.482 & 0.649 & 0.772 & 0.866 \\ \cline{2-7} 
 & 16384 & 0.321 & 0.504 & 0.657 & 0.786 & 0.869 \\ \hline
\multirow{3}{*}{$L_{seg}$+$L_{fft}$} & 4096 & 0.543 & 0.766 & 0.953 & 0.982 & 0.999\\ \cline{2-7} 
 & 8192 &0.551 & 0.788 & 0.956 & 0.984 & 0.999 \\ \cline{2-7} 
 & 16384 & 0.597 & 0.828 & 0.947 & 0.978 & 0.999\\ \hline
\end{tabular}
\label{ablative_loss_components}
\end{table}

To calculate the probability of correct detection ($P_{\text{d}}$) and probability of false alarm ($P_{\text{fa}}$) using an energy detector we use the radar parameters specified in \cite{Donoughue2019}, where we obtain 52.86\% $P_{\text{d}}$ and   8.47\%$P_{\text{fa}}$. For our model we record a 98.36\% $P_{\text{d}}$ and   1.34 \%$P_{\text{fa}}$.

\section{Conclusion and Future Directions}
This paper presents a novel segmentation model for detecting a signal and localising it in the time domain. The model we propose can also be used for estimating the FFT representation of the signal, which helps in limiting the search scope for downstream signal processing tasks for electronic support. We evaluate our models on a synthetic dataset which represents a EW environment. In future work, the dataset can be extended to incorporate more challenging and realistic channel effects and other segmentation models including multi-stage models can be explored. 
\section{Acknowledgement}
The research for this paper received funding support from the Queensland Government through Trusted Autonomous Systems (TAS), a Defence Cooperative Research Centre funded through the Commonwealth Next Generation Technologies Fund and the Queensland Government.

\vspace{12pt}
\color{red}

\end{document}